\documentclass[%
reprint,
superscriptaddress,
 amsmath,amssymb,
 aps,
 pra,
]{revtex4-1}

\usepackage{graphicx}
\usepackage{dcolumn}
\usepackage{bm}
\usepackage[mathlines]{lineno}

\begin{document}

\title{Testing conformal mapping with kitchen aluminum foil}

\author{S.~Haas}
\affiliation{Institute for Particle Physics, ETH Zurich, Switzerland}

\author{D.~A. Cooke}
\affiliation{Institute for Particle Physics, ETH Zurich, Switzerland}

\author{P.~Crivelli}
\email[]{crivelli@phys.ethz.ch}
\affiliation{Institute for Particle Physics, ETH Zurich, Switzerland}

\begin{abstract}

We report an experimental verification of conformal mapping with kitchen aluminum foil. This experiment can be reproduced in any laboratory by undergraduate students and it is therefore an ideal experiment to introduce the concept of conformal mapping. The original problem was the distribution of the electric potential in a very long plate. The correct theoretical prediction was recently derived by A. Czarnecki \cite{andrzej}.

\end{abstract}

\maketitle

\section{Statement of the problem}\label{sec:Problem}

\begin{figure}[h!]
   \centering
   \includegraphics[scale=0.3]{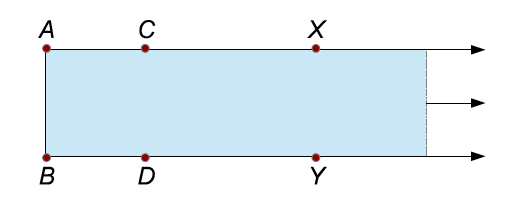}
   \caption{A semi-infinite metal ruler to which the voltage $U_s$ is applied between A and B. Between C and D the voltage  $U_d$ is being measured.}
   \label{fig:ruler}
\end{figure}

The initial problem is part of the collection of the Moscow Institute of Physics and Technology \cite{cahn}. A voltage $U_s$ is applied to the corners A and B of a semi-infinite long metallic ruler as shown in Fig. \ref{fig:ruler}).

The question is: when one measures the voltage difference $U_d$ between C and D, how does the voltage difference behave further down the ruler? Assuming that the thickness of the ruler can be neglected, this can be reduced to a two-dimensional problem. A. Czarnecki derived the correct solution to this problem conformal mapping \cite{andrzej}. In this tutorial, we present the experimental verification of these new calculations.

\section{Conformal mapping}

Conformal mapping is a mathematical technique which is widely used not only in physics
but also in engineering. The main idea behind this technique is to simplify a certain problem
by mapping it to a better suited geometry in order to simplify its solution. 

\subsection{Mathematical definition}
A complex function $f : U \to \mathbb{C}$ is called holomorphic, if it is complex differentiable at any point in its domain. Or in other words, if the following limit exists:
\begin{equation}
f'(z_0) = \lim_{z \to z_0}\frac{f (z) -f (z_0)} {z - z_0}
\end{equation}
A holomorphic function $g: U \to \mathbb{C}$ is said to be conformal if $g'(z) \neq 0 \, \forall  z \in U$. Conformal functions have the properties to preserve locally angles and the shapes of infinitesimally small figures.

\subsection{Example: the mercator projection}
\begin{figure}[h!]   
\centering
   \includegraphics[scale=0.1]{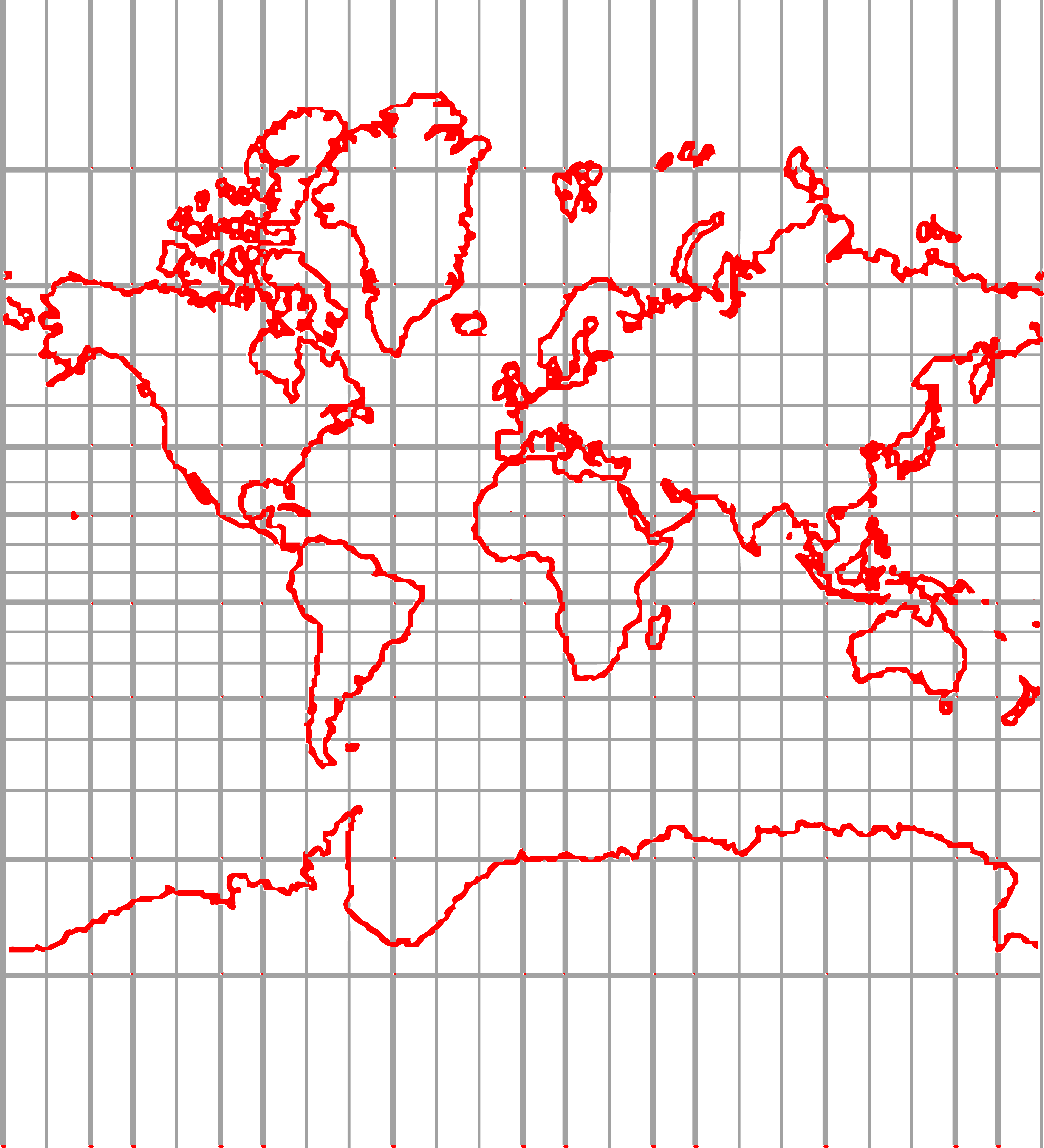}
   \caption{Mercator projection \cite{mercator}.}
   \label{fig:mercator}
\end{figure}

The Mercator projection is a cylindrical map projection. It is probably the most common way to map the spherical earth surface in two dimension. The corresponding map is:
\begin{equation}
x = R (\theta -\theta_0)
\end{equation}
\begin{equation}
y = R \ln\Big[\tan{\frac{1}{4}\pi-\frac{1}{2}\phi}\Big]
\end{equation}
Where $R$ is the Earth radius, $\phi$ is the latitude, $\theta$ the longitude and $\theta_0$ an arbitrary central meridian (commonly chosen to be the one of Greenwich). This mapping satisfies the above condition of a conformal map and visualises well its properties. The circles of longitude
and latitude are perpendicular on the map and on small scales the shapes of objects are preserved. whereas large objects can change their shape and size depending where they are located on the globe. For example according to Fig. \ref{fig:mercator} the size of Greenland and Africa would be of the same order.

\section{Solution}

The solution of the problem presented in Sec. \ref{sec:Problem} as suggested in \cite{cahn} starts with comparing the potential differences $U_x$ at two nearby pairs of points:
\begin{equation}
U_{x+dx}-U_x=\alpha(x)U_xdx
\end{equation}
It is then argued that since the ruler is semi-finite, the coefficient $\alpha(x)=\alpha$ is constant and does not depend on position. This leads to the differential equation: 
\begin{equation}
U_x'=\alpha U_x
\end{equation}
If this equation would hold everywhere one would get the final solution:
\begin{equation}\label{eq:originalSol}
U_x=U_s\left(\frac{U_s}{U_d}\right)^{-x/d},
\end{equation}
where $U_s$ is the voltage applied at the edge and $U_d$ the measured voltage difference at a given point with a distance $d$ from the edge. However, the ruler is only semi-infinite and not infinite, so the assumption that the voltage does not depend on the position is not fulfilled. In fact, points very close to the beginning of the ruler do not have the same neighbourhood as points further down the ruler.\par

Czarnecki \cite{andrzej} derived the correct solution to this problem using conformal mapping. Complex coordinates $z=x+iy$ were introduced such that the corner B corresponds to $z_B=0$ and the upper corner A to $z_A=i$. Looking at the image of the ruler under the mapping $z \to w(z)=e^{\pi z}$, one sees that the corners A and B are mapped on the $x$-axis, $w_A=(-1,0)$ and $w_B=(1,0)$ as shown in Fig. 2. 

\begin{figure}[h!]   
   \centering
   \includegraphics[scale=0.16]{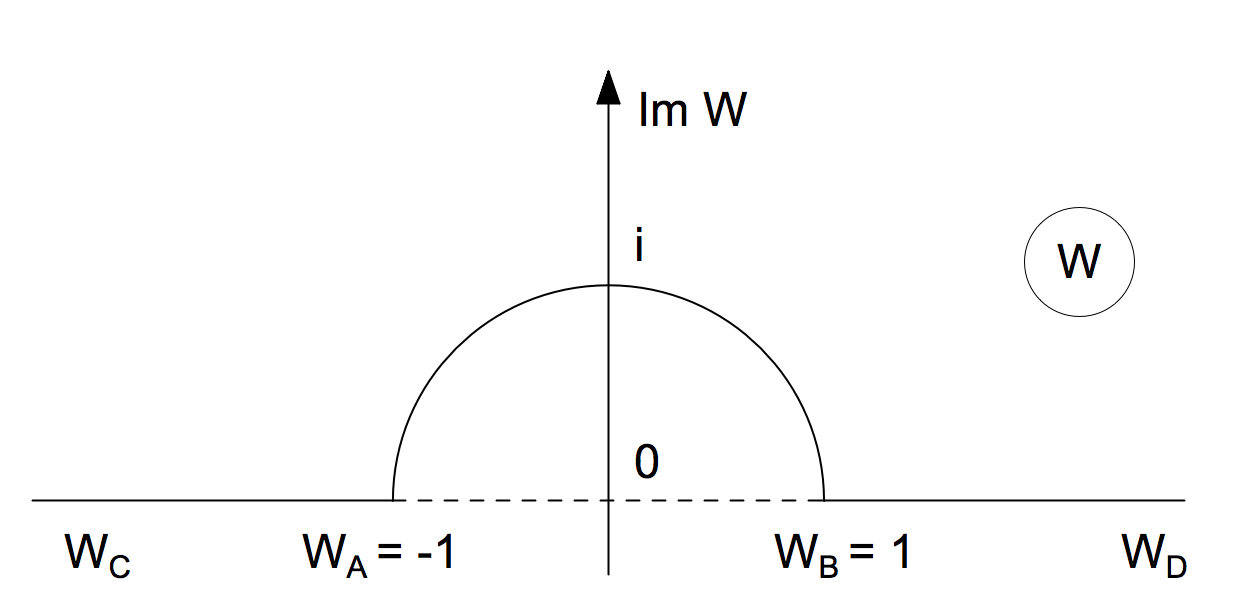}
   \caption{Image of the ruler under the conformal mapping $w(z)=e^{\pi z}$}
\end{figure}
If an infinite ruler would be mapped with this function, this would cover the whole upper half-plane. The actual advantage of this mapping is that every function that only depends on the distance to the corners is now symmetric to the real axis since both corners lie on that axis. This symmetry leads to the solution: 
\begin{equation}\label{eq:AndrzejSol}
U(x)=\frac{U_s}{c-\ln s}\ln \frac{1+e^{-\pi x}}{1-e^{-\pi x}},
\end{equation}
where $s$ is the length of the contacts and $c$ is a constant depending on their detailed geometry but not on their size if $s$ is small. The expression $(c-\ln s)$ describes physical rather than idealized contacts and we thus refer to it as the reality factor. These contacts parameters are dependent on the width of the ruler.\par

If one assumes that the distance between the left corner and the position where we measure the voltage difference is sufficiently large (more then a third of the width of the stripe), this formula can be approximated to:
\begin{equation}
U(x)=U_{d_1} \left( \frac{U_{d_2}} {U_{d_1}} \right)^{\frac{x-d_1}{d_2-d_1}}
\end{equation}
where $U_{d_1}$ and $U_{d_2}$ are the voltage differences measured at distances $d_1$ and $d_2$ from the edge.


\section{Experimental verification}

\subsection{Experimental setup}
\begin{figure}[h!]   
   \centering
   \includegraphics[scale=0.2]{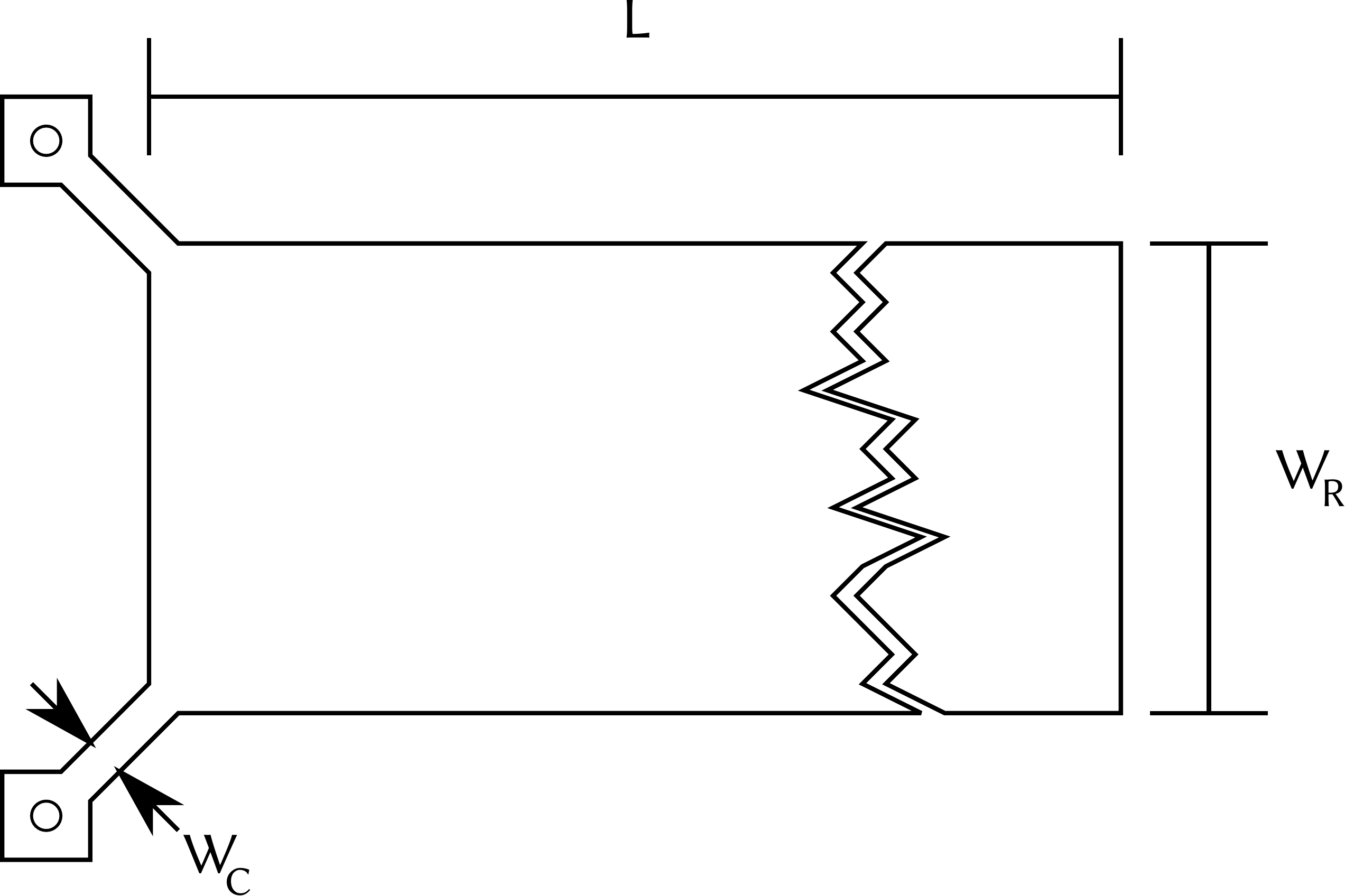}
   \caption{Sketch of the experimental setup.}
   	\label{setup}
\end{figure}

To realize this experiment the following basic laboratory equipment is required:
\begin{itemize}
  \item Standard power supply (30 V, 3 A)
  \item Standard voltmeter ($\pm0.01$ mV)
  \item Aluminium foil (10-50 $\mu$m, typical thickness of aluminium kitchen foil)
\end{itemize}
The experiment consists of applying and measuring the voltage difference at different points. To simulate a semi-infinite metallic ruler, the metal stripe made of aluminium foil was cut much longer than actually needed. The wires, used to apply the voltage difference, were pulled through holes in the contact stripes as shown in Fig. \ref{setup}. The experimental configuration can be described by the following parameters: the length of the ruler $L$, its width $W_R$ , its thickness $T$ and the width of the contact stripes  $W_C$ (see Fig. \ref{setup}).

\subsection{Results}

The voltage difference was measured in 5 mm steps form 0 to 50 mm. These measurements were performed with different settings in order to investigate the influence of the following factors: $L$,  $T$, $W_R$ and $W_C$. The experimental uncertainties to be taken into account are the ones of the voltmeter ($\pm$0.01 mV) and the one of the measuring position ($\pm$0.3 mm).

\begin{figure}[h!]   
   \centering
   \includegraphics[scale=0.4]{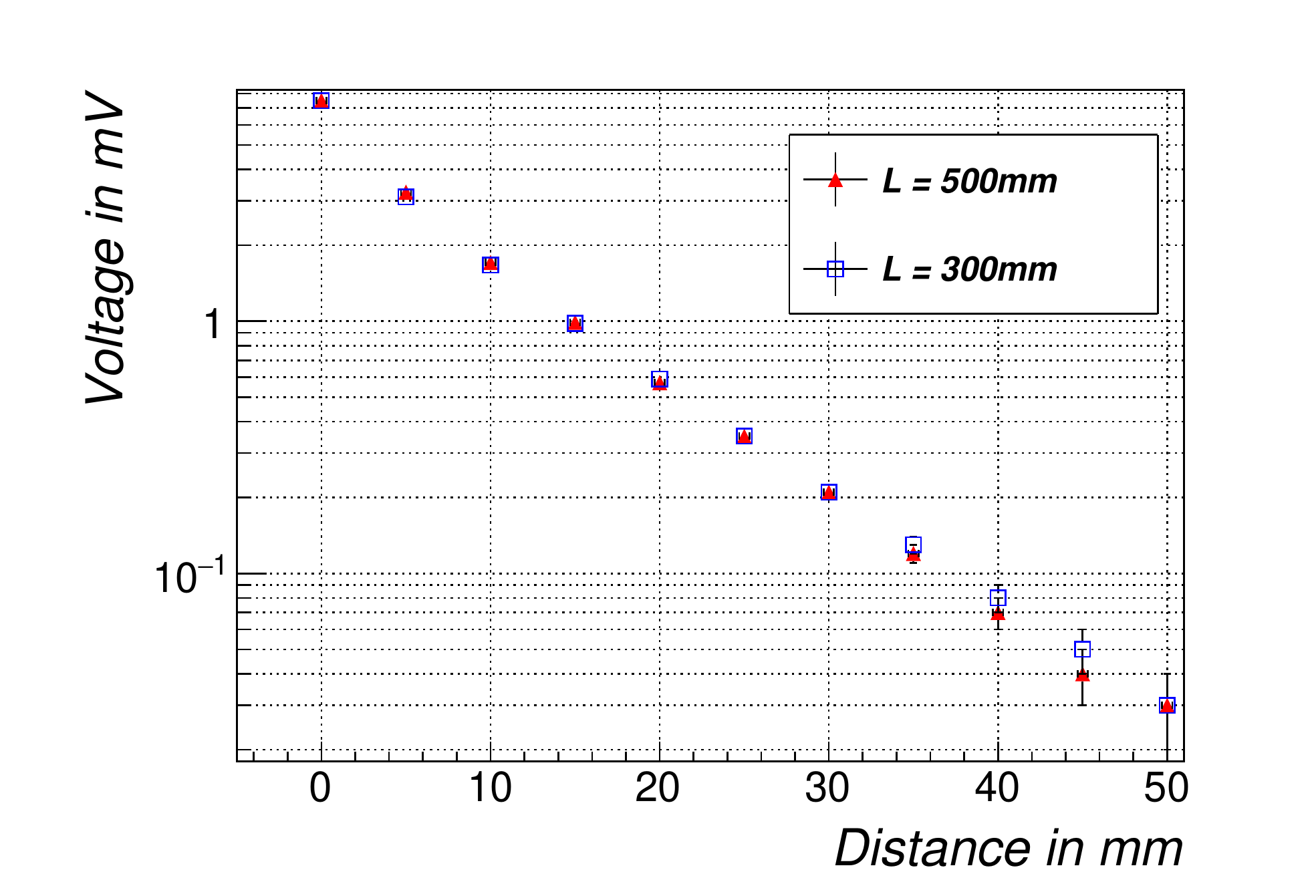}
   \caption{Influence of the foil length $L$ on the voltage difference.}
   \label{VvsL}
\end{figure}

\begin{figure}[h!]   
   \centering
   \includegraphics[scale=0.4]{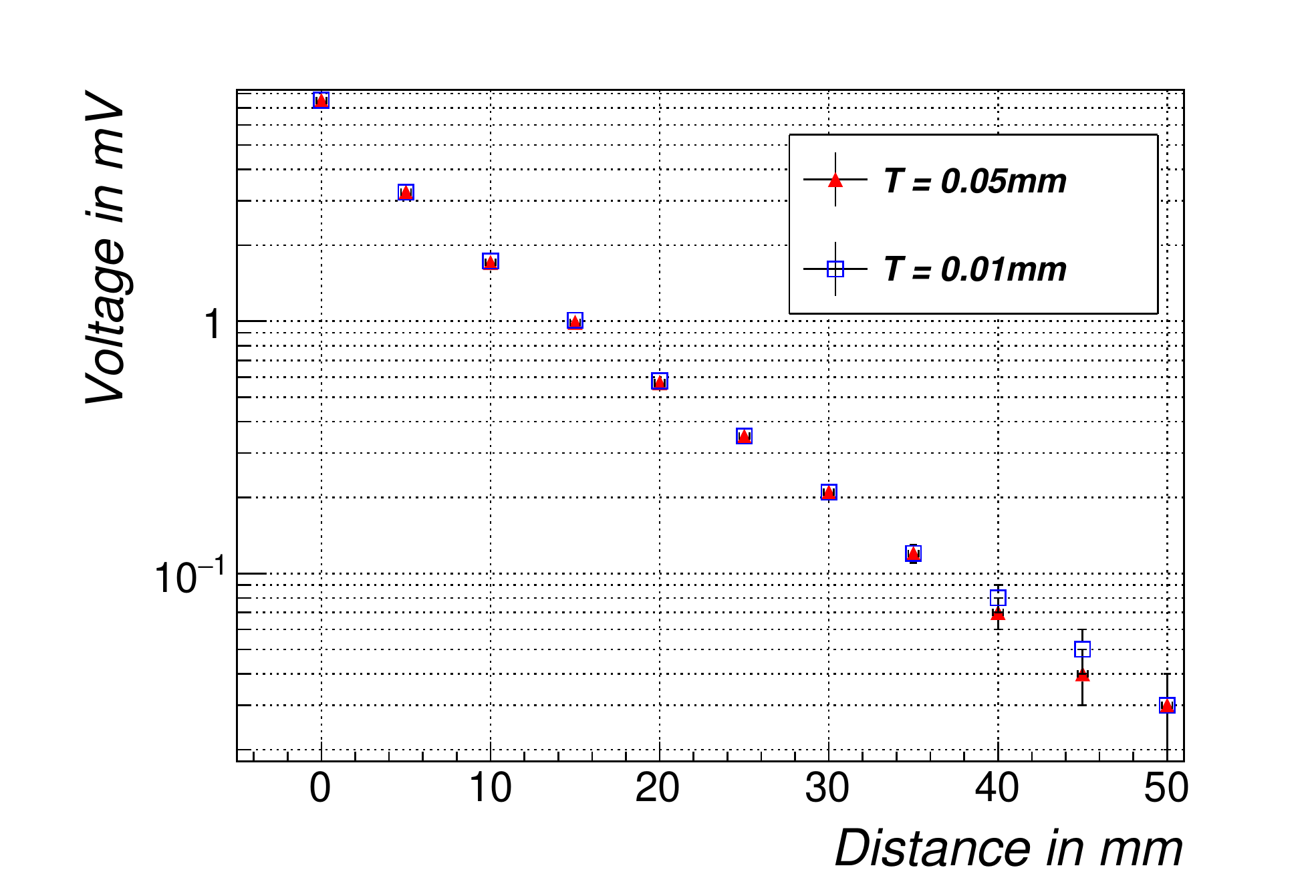}
   \caption{Influence of the foil thickness $T$ on the voltage difference}
   \label{VvsT}
\end{figure}

The results for $L$ and $T$ are presented in Figs. \ref{VvsL}-\ref{VvsT}. As one can see, the measured points for ruler lengths of 300 and 500 mm are the same within the experimental errors, one can thus conclude that a ruler with more than 300 mm is sufficiently long for our experiment and it is a good approximation of a semi infinite ruler. The results are also unaffected when using two different thicknesses of $T$=0.01 and 0.05 mm. Therefore aluminium kitchen foil is thin enough to approach a 2-dimensional problem as required by this experiment. 

\begin{figure}[h!]   
   \centering
   \includegraphics[scale=0.4]{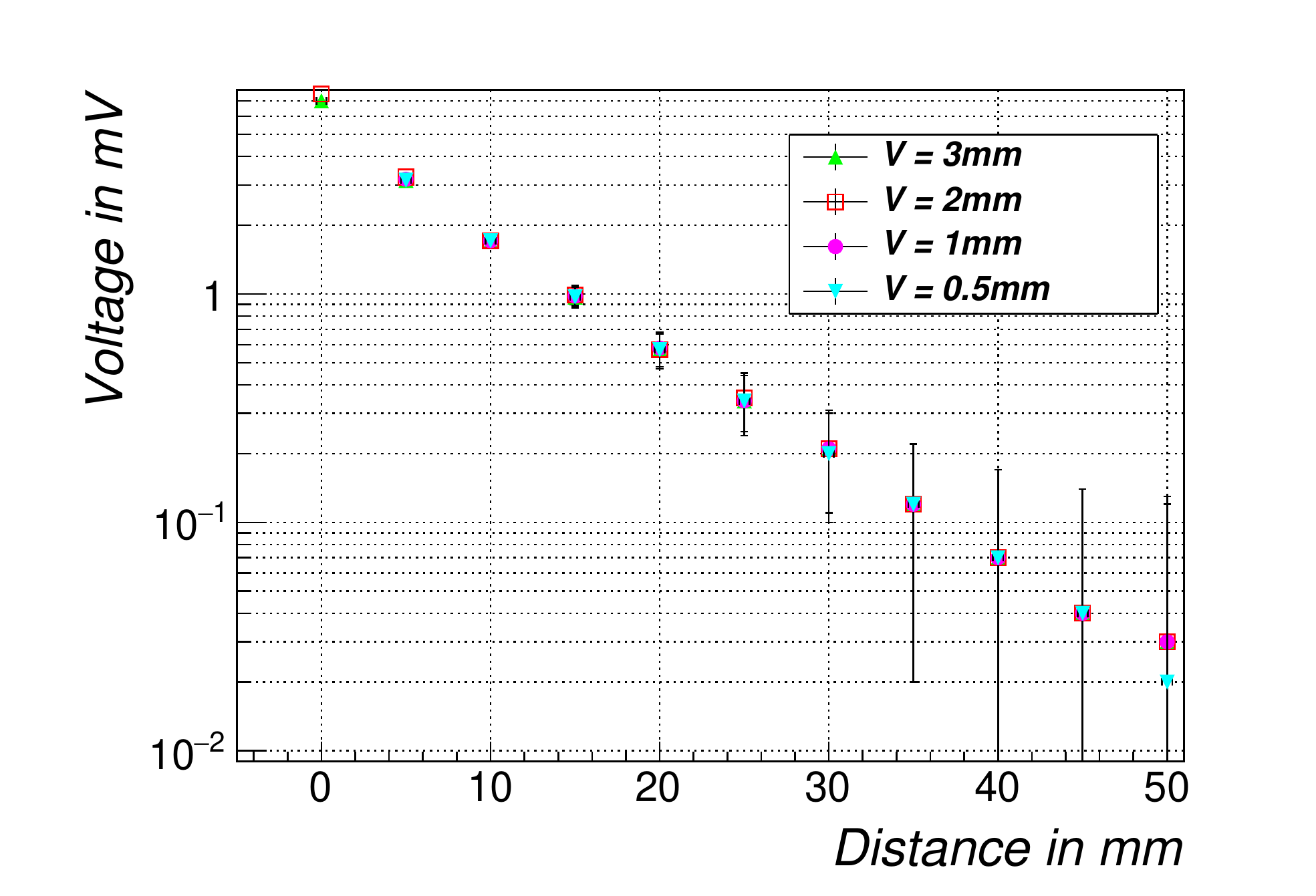}
   \caption{Dependence of the  width of the voltage stripe $W_c$ on the voltage difference.}
      \label{VvsWC}
\end{figure}

To study the effect of the contacts geometry and thus the $(c-\ln s)$  parameter of Eq.\ref{eq:AndrzejSol}, the width $W_C$ was varied. Apart from the first measured point at $x=0$ mm, the obtained values are the same for all the others distances within the experimental errors (see Fig. \ref{VvsWC}). 

\begin{figure}[h!]   
   \centering
   \includegraphics[scale=0.4]{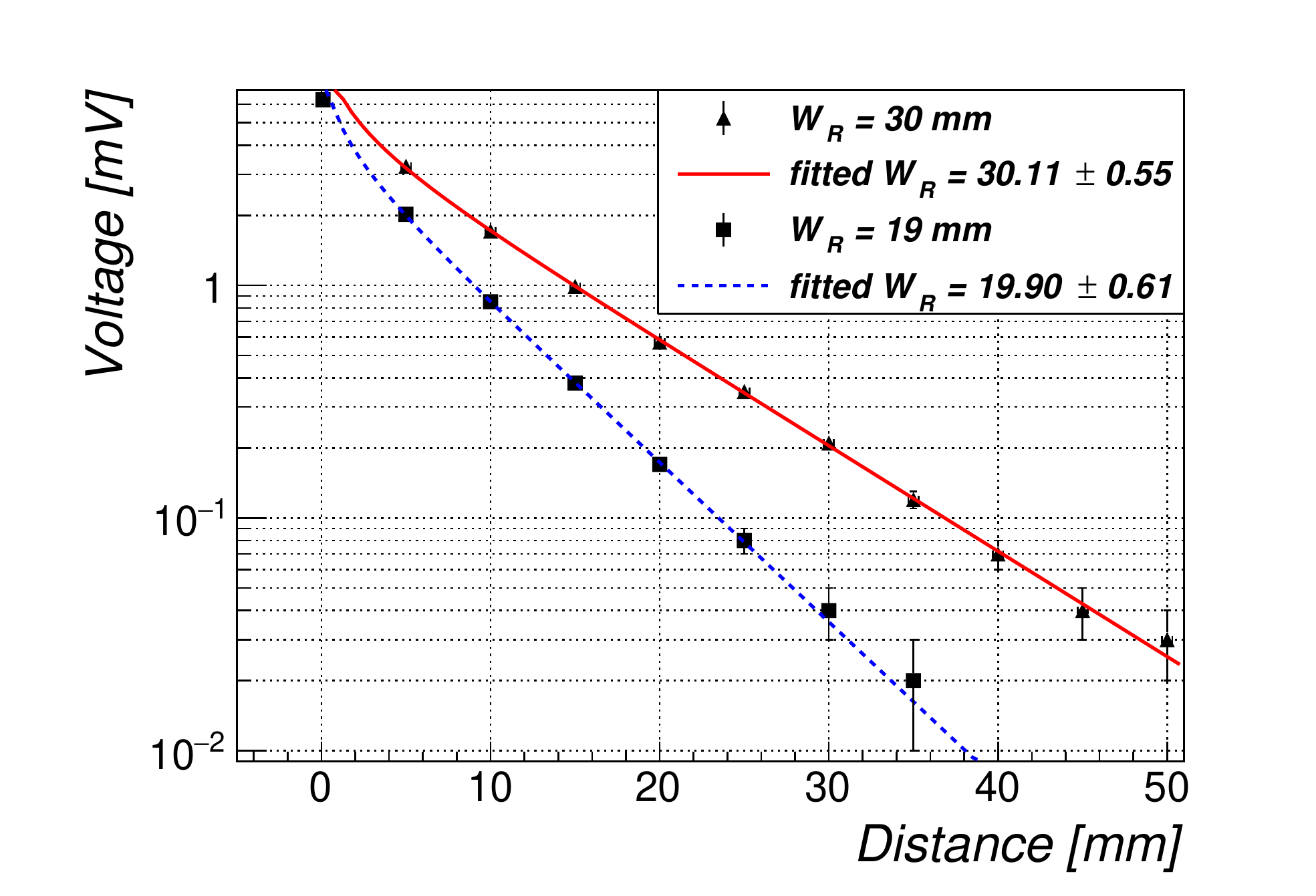}
   \caption{Influence of the width $W_R$ on the voltage difference and fit to the data using Eq. \ref{eq:AndrzejSol}}
   \label{WR}
\end{figure}

The last parameter to be investigated is the influence of the width $W_R$ of the ruler. According to the original calculation (Eq. \ref{eq:originalSol}) one would expect that the measured values are independent on $W_R$. This is in contradiction with the data as shown in Fig. \ref{WR} confirming the inadequacy of this solution and correctness of the newer calculations 
 
\section{Conclusions}
Our measurements are in very good agreement with the predictions of the new calculations obtained applying conformal mapping by Czarnecki and point out that the original solution in which the problem of calculating the distribution of the electric potential in a very long plate was proposed is not adequate. This problem is a very nice example of the application of conformal mapping and since this experiment is very simple and uses only basic equipment, it can reproduced in any undergraduate laboratory thus providing a very good introduction to students on this subject.



\end{document}